\documentclass[a4paper,fleqn,usenatbib]{mnras}

\usepackage{newtxtext,newtxmath}

\usepackage[T1]{fontenc}
\usepackage{ae,aecompl}

\usepackage{graphicx} 
\usepackage{amsmath} 
\usepackage{amssymb} 

\title[KH instability in the Orion nebula]{The Kelvin-Helmholtz instability in the Orion nebula: The effect
of radiation pressure }

\author[Yaghouti et al.]{
Akram Yaghouti,$^{1}$
Mohsen Nejad-Asghar,$^{2}$
Shahram Abbassi$^{1,3}$\thanks{E-mail: abbassi@um.ac.ir}
\\
$^{1}$Department of Physics, School of Science, Ferdowsi University of Mashhad, Mashhad,PO Box 91775-1436 Iran\\
$^{2}$Department of Physics, University
of Mazandaran, Babolsar, Iran\\
$^{3}$School of Astronomy , Institute for Studies in Theoretical Physics and Mathematics, PO Box 19395-5531, Tehran, Iran
}

\date{Accepted XXX. Received YYY; in original form ZZZ}

\pubyear{2015}

\begin{document}
\label{firstpage}
\pagerange{\pageref{firstpage}--\pageref{lastpage}}
\maketitle

\begin{abstract}
  The recent observations of rippled structures on the surface of the Orion molecular cloud (Bern\'{e} et al. 2010), have been attributed to the Kelvin-Helmholtz (KH) instability. The wavelike structures which have mainly seen near star-forming regions taking place at the interface between the hot diffuse gas, which is ionized by massive stars, and the cold dense molecular clouds. The radiation pressure of massive stars and stellar clusters is one of the important issues that has been considered frequently in the dynamics of clouds.
Here, we investigate the influence of radiation pressure, from well-known Trapezium cluster in the Orion nebula, on the evolution of KH instability. The stability of the interface between HII region and molecular clouds in the presence of the radiation pressure, has been studied using the linear perturbation analysis for the certain range of the wavelengths. The linear analysis show that consideration of the radiation pressure intensifies the growth rate of KH modes and consequently decreases the e-fold time-scale of the instability. On the other hand the domain of the instability is extended and includes the more wavelengths, consisting of smaller ones rather than the case when the effect of the radiation pressure is not considered. Our results shows that for $\lambda_{\rm KH}>0.15\rm pc$, the growth rate of KH instability dose not depend to the radiation pressure. Based on our results, the radiation pressure is a triggering mechanism in development of the KH instability and subsequently formation of turbulent sub-structures in the molecular clouds near massive stars. The role of magnetic fields in the presence of the radiation pressure is also investigated and it is resulted that the magnetic field suppresses the effects induced by the radiation pressure.  
\end{abstract}



\section{Introduction}
The wave-like structures at the surface of the Orion nebula has been
recently observed by Bern\'{e}, Marcelino and Cernichora~(2010),
mainly seen near massive star-forming regions (Figure 1). They proposed that
these ripples may be produced by some hydrodynamical instabilities
taking place in the interface between the hot diffuse gas, which is
ionized by massive young stars, and the cold dense molecular cloud,
which was nursery of the star-forming pawns. The rippled structures
are difficult to be detected in the interstellar medium, but it is
believed that they are relatively widespread in the interface of
different regions. For example, Sahai, Morris and Claussen~(2012)
reported observations of source IRAS 20324+4057 which show an
extended, limb-brightened, tadpole-shaped nebula with ripples in the
tail of the tadpole. Also, Proplyds, which are ionized
protoplanetary disks around young stars (e.g., Kim et al. 2016) may
reveal some rippled structure in the interface of the diffused ionized
gas of disk and the dense parental molecular cloud.

The most likely hydrodynamical instability that may be able to
describe the ripples appropriately, seems to be the Kelvin-Helmholtz
(KH) instability. There are some significant evidences which confirm
this claim that rippled structures has been created by KH
instability: the molecular cloud is subject to an important velocity
gradient that is due to champagne flow occurred by the exploding HII
region, the passage of flow of diffuse HII gas, and the acceleration
of the molecular cloud by it that makes a velocity shear in the interface between two regions. In this way, Bern\'{e} and Matsumoto~(2012)
accomplished a linear MHD analysis applied to the situation of the
Orion's ripples, using physical parameters determined
observationally. They studied stability of boundary layer between
the HII region and molecular cloud, and obtained the growth rate of
instability. They indicated that the wavelengths of rippled
structures are compatible to that obtained theoretically and
therefore it confirms that these ripples are arisen by KH
instability. They also claimed that KH instability plays an
important role in the formation of small scale turbulent sub-structures near the star-forming regions, and can be considered a relevant process for mixing of the
chemical elements inside HII region with molecular clouds in regions
where planetary systems are formed around young stars. Hendrix,
Keppens and Camps~(2015) used the numerical modelling to investigate occurrence of the
ripples in the Orion nebula. They considered coupling of gas and dust
dynamics and radiative transfer on the KH instability, in a way that the observations were directly
compared to the modelling results. They concluded that the formation
of KH instability is not inhibited by addition of dust, while, to
get agreement with the observed KH ripples, the assumed geometry
between the background radiation, the billows, and the observer, was
seen to be of critical importance.

The radiation pressure of massive stars and stellar clusters is one of the issues that has been considered frequently in the
dynamics of clouds. For example, the radiation pressure is a
physical process that can disrupt giant molecular clouds, and has
been mentioned as an important feedback mechanism which can affects
the efficiency of star formation (e.g., Murray, Quataert and
Thompson~2010; Krumholz and Tan~2007; Kennicutt~1998). Also, the exerted radiation pressure of ambient non-ionizing radiation on the cloud cores and filaments, can alter
the structure of them and can promote the gravitational collapse
(e.g., Seo and Youdin~2016).

The Trapezium cluster is an illustrious source and responsible for much
of the illumination of the surrounding Orion nebula. It is a tight
group of stars in the center of the Orion nebula, and includes
several extremely bright young stars that the luminosity of each one
of them is several thousand times of that of the Sun
(Sim\'{o}n-D\'{\i}az et al. 2006). In this way, we expect that the
radiation pressure of the Trapezium cluster, which are in the vicinity
of Orion's ripples, may have significant effects on development of
the rippled structure. For this
purpose, we investigate the effect of radiation pressure of
the Trapezium cluster on development of the KH instability in the different
situations of the Orion nebula. In section~2, we do estimations
for measurement of the radiation pressure of the Trapezium cluster. In
section~3, we represent an equilibrium model for the interface layer
between the HII region and molecular cloud. In section~4, we perturb the
equations and then solve them numerically with introducing the
suitable boundary conditions. The numerical results are given in
section~5, and the regions with the strong magnetic fields are discussed
in section~6. Finally, section~7 is devoted to the discussion.
\section{ESTIMATIONS}
The Trapezium cluster of the Orion nebula is a group of stars of spectral types
O and B, with large values of the rate of emission and luminosity of
non-ionizing and ionizing photons. We consider Trapezium as a compact cluster of OB stars that might be treated as a source with certain luminosity $L^T$. We can roughly estimate the radiation pressure of this association at the place of the rippled structures. This source has total luminosity $L^T = L_{\rm n} +
L_{\rm i}$ where $L_{ \rm n}$ and $ L_{\rm i}$ are the non-ionizing
and ionizing luminosities of the Trapezium cluster for $h\nu <13.6 $ eV 
and $h\nu > 13.6 $ eV  photons, respectively. For massive
stars, and the clusters including massive stars whose luminosity comes mostly from ionizing photons, we have approximately $L^T = L_{\rm i}$. The radiation from the Trapezium cluster spreads over a surface area of the sphere whose radius is the distance from the Trapezium's center to the interface between the HII region and molecular cloud (Figure 2). The total average flux of radiation at the distance $d$ is $ F^T_{ rad} = L^T/4\pi d^{2} $. Only a portion of emitted radiation from Trapezium, strikes the layer and some fraction of photons may be absorbed or dispersed via scattering in the HII region. If we consider only $10\%$ of photons which enter to the interface layer between two regions of HII and molecular cloud
 and they transfer momentum and impose the
radiation pressure on it, then the radiation pressure on
layer can be approximately estimated by the flux of radiation as
$P_{ r}\approx 0.1 \frac{ F^T_{ rad}}{c}$, where c is the
speed of light. If we express $L^T$ in unit of solar luminosity
($\rm L_\odot$) and $d$ in the scale of parsec, we have
\begin{eqnarray}\label{eqn1}
P_{ r} \approx 1.1\times10^{-16}\frac{
L^T_{(L_{\odot})}}{d_{(pc)}^{2}}~\rm dyn~cm^{-2}.
\end{eqnarray}
\begin{figure}
\centering
\includegraphics[width=7.5cm,height=5cm]{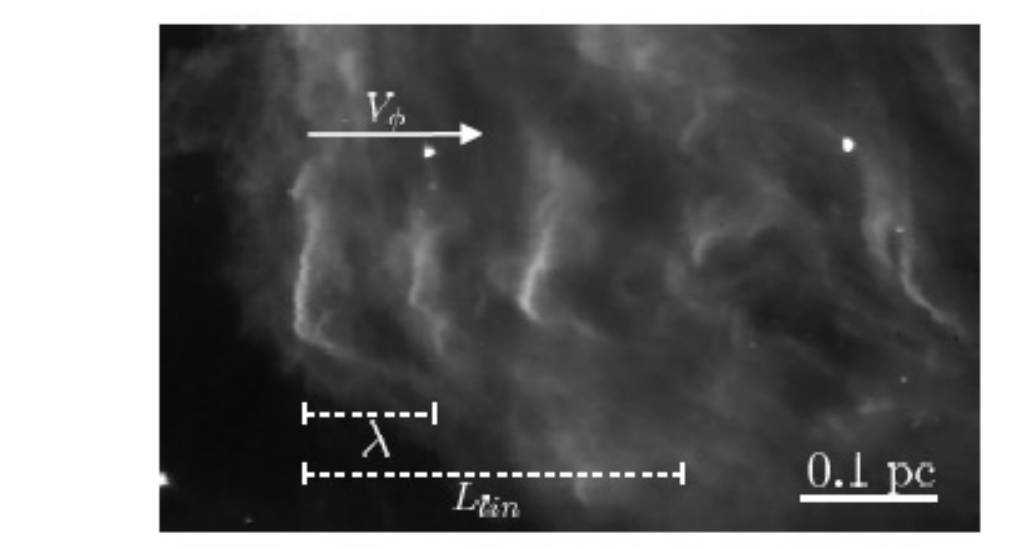}
\caption{ Observation of the ripples in the Orion nebula at 8 $\mu m$, taken with the Spitzer Infrared Array Camera. $ \lambda $ denote the spatial wavelength of the structure, $ L_{\rm lin} $ is the distance over which the instability is linear and and $V _{\phi}$ is the phase velocity of KH waves at the surface of the cloud. This figure is from Bern\'{e} \& Matsumoto (2012). \label{fig1}}
\end{figure}
\begin{figure}
\centering
\includegraphics[width=4.3cm,height=5.2cm]{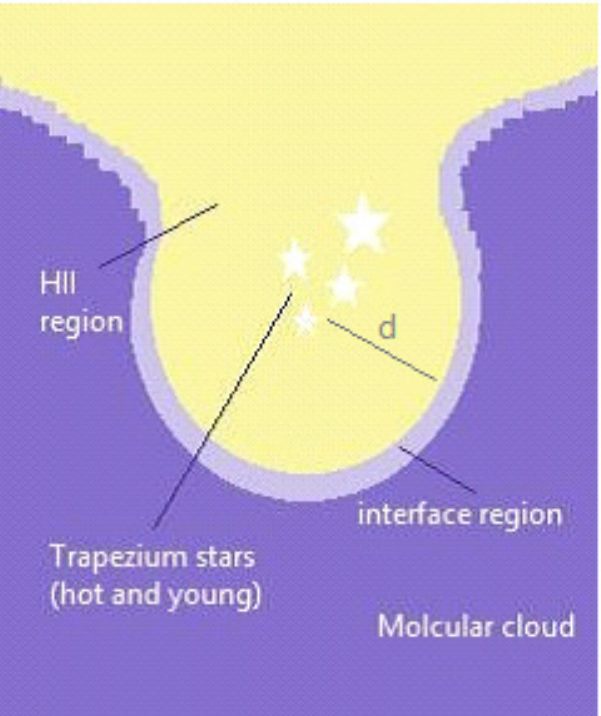}
\caption{Schematic picture of interface between the HII region and molecular cloud in the presence of radiation field of the Trapezium cluster at the approximate distance $d$.\label{fig1}}
\end{figure}
Considering the luminosity of OB stars as $L _{\rm OB} = 0.8\times 10^{39} \rm erg s^{-1} \simeq 2\times 10 ^{5} L_ {\odot}$ (Drain 2011), and using the analyzed results of Sim\'{o}n-D\'{\i}az et al. (2006) by including about 10-100 stars in the cluster, we predict that the total luminosity of this stellar
association is approximately between $10^{6}$ to $10^{7} L_{\odot} $. For a
typical distance of 2pc, according to the scale of distance in the
Figure~1 of Bern\'{e}, Marcelino and Cernichora~(2010), the
radiation pressure at the interface layer can be estimated as
$2.7 \times 10^{-11} \rm dyn~cm^{-2} \lesssim 
P_{ r} \lesssim 2.7\times 10^{-10}\rm dyn~ cm^{-2}$. On the other hand, if we adopt the observational data for
the density and temperature of the HII region and the molecular cloud, which
were used by Bern\'{e} and Matsumoto~(2012): $n_{ II}=20
\rm cm^{-3}$, $T_{ II }=10^4\rm K$, $n_{ MC}=10^{4}\rm cm^{-3}$,
$T_{ MC}=20\rm K$, the gas pressure at the interface is equal to
$ P_{ g}= k_B n_{ II} T_{ II} = k_B n_{MC} T_{MC} =
2.7\times10^{-11} \rm dyn~ cm^{-2}$. Thus, by comparing
the gas pressure with the estimated radiation pressure, it can be
deduced that $1 P_{ g} \lesssim P_{r} \lesssim 10 
P_{g}$. For a distance of 3pc, we obtain this estimation as
$0.3 P_{g} \lesssim P_{r}\lesssim 3 P_{g}$. If we consider the more numbers of photons $(>10\%)$ to enter to the layer and transfer momentum, then the more amounts of radiation pressure are obtained. In the subsequent sections, we consider the maximum allowed value of $\Pi \equiv P_{r}/P_{g}$ equal to $100$. 
With these estimations, it is expected that the radiation pressure may have a significant dynamical impact on the formation of rippled structures of the Orion molecular cloud around the HII region, specially on the instabilities which are constituted there. For this purpose, we
investigate the influence of radiation pressure on the growth rates
of the KH instability for real physical conditions in the Orion nebula. We want to obtain the corresponding time-scales of these KH instabilities, by using the linear perturbation analysis method.
\section{THE EQUILIBRIUM MODEL FOR RADIATED LAYER}
\begin{figure}
\includegraphics[width=\columnwidth]{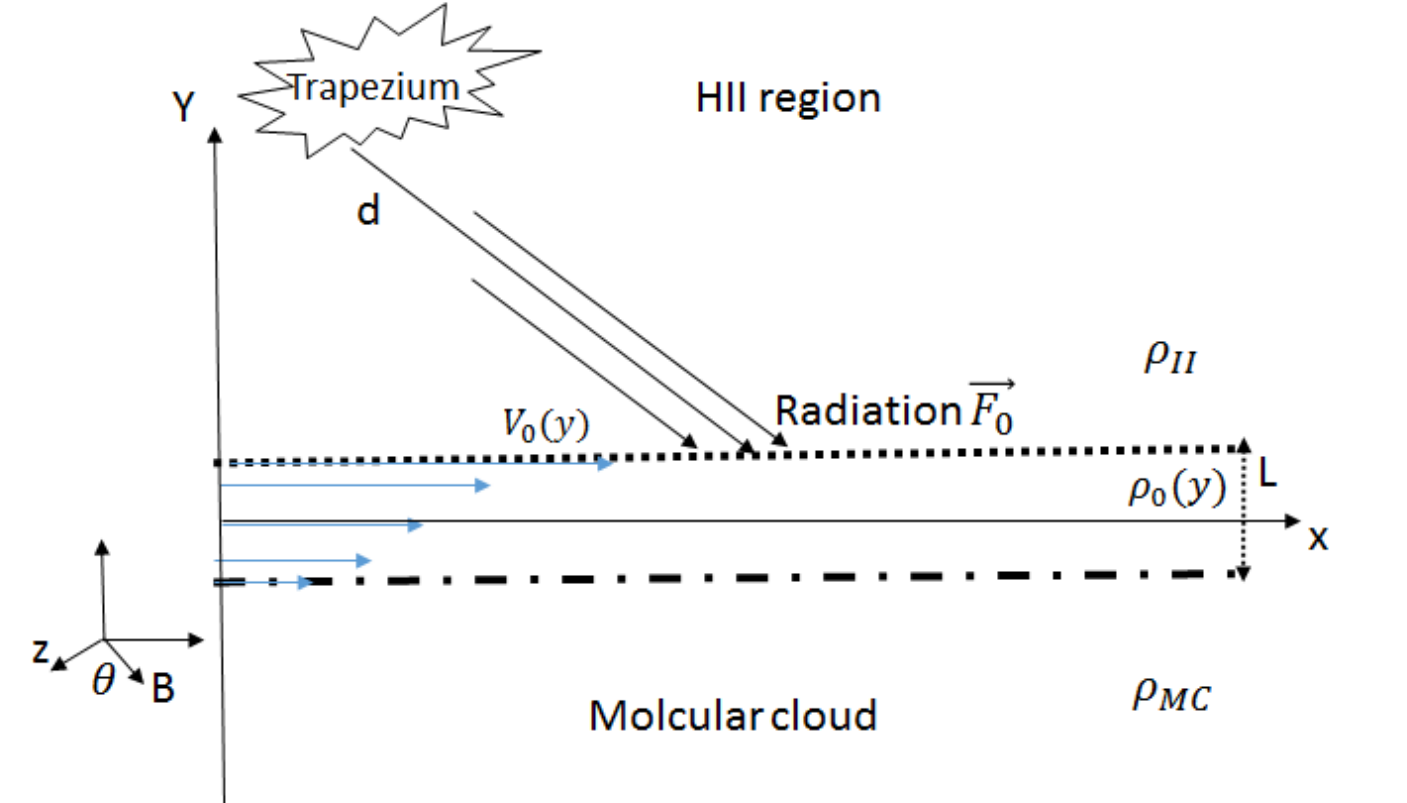}
\caption{Equilibrium model for the interface layer between the HII region and molecular cloud in the presence of radiation field of the Trapezium cluster.\label{fig2}}
\end{figure}

Here, we investigate the equilibrium of the interface layer between
HII region and molecular cloud, in the presence of the radiation
field. The geometry used for the interface layer is a two dimensional set-up that is
depicted in the Fig.~\ref{fig2}. The upper part corresponds  to the hot,
low density HII region with density and temperature $ \rho_{ II}$
and $ T_{ II}$, respectively; the lower part represents the cold,
high density molecular cloud with density and temperature
$\rho_{MC}$ and $ T_{MC}$, respectively; and both are separated
by a thin middle layer with thickness $L$, which is estimated about 0.01pc by means
of observations (Berne, Marcelino and Cernichora~2010). The radiation pressure acts on the matter of this interface layer.

The coordinates are chosen so that the interface layer is in the
$x$-axis direction; the density and velocity gradients are along the $y$-axis.
Here, we use the hyperbolic-tangent form,
for the profiles of the velocity (Miura and Pritchett 1982) and density across the transition layer between HII region and molecular cloud, as follows
\begin{eqnarray}\label{8}
{\bf V}_0(y)=\frac{V_{0}}{2}\left(1+\tanh\frac{y}{L}\right){\bf
e}_{x},
\end{eqnarray}
\begin{eqnarray}\label{9}
\rho_{0}(y)=\frac{\rho_{
II}}{2}\left[(1-\alpha)\tanh\frac{y}{L}+(1+\alpha)\right]
\end{eqnarray}
where $V_{0}$ is the total variation of velocity across the
transition layer, ${\bf e}_{x}$ is the unit vector along the
$x$-axis, and $\alpha \equiv \rho_{MC} / \rho_{ II}$.
The magnetic field is assumed to be in the $xz$-plane and its
orientation is defined by the angle $\theta $ between $\bf B$ and
$z-$ axis. In general, we have ${\bf B}_0 = B_{0}(y) \sin\theta {\bf e}_x
+B_{0}(y) \cos\theta {\bf e}_z $ so that the magnetic divergence
equation, $\nabla \cdot {\bf B}_0 =0$ is satisfied. We choose the
density and temperature of HII region and molecular cloud so that
the gas pressure, $ P_{0g}$, be equal between the upper and lower
mediums and is independent of $y$. In this way, the general
equilibrium condition of the interface layer, in the $y$ direction,
\begin{eqnarray}
P_{0g}+P_{0r}(y)+\frac{B_{0z}^{2}(y)}{2\mu_0}= const.,
\end{eqnarray}
implies that the magnetic field at the HII region must be smaller than
its value at the molecular cloud because the radiation pressure at
the HII region is greater than its value at the molecular cloud. This
result is in consistent with the observations of Abel et al.~(2004)
and Brogan et al.~(2005) who found that the magnetic field varies
between 5 nT at the Trapezium region to 25 nT at the
Veil region.

\section{BASIC EQUATIONs AND LINEAR PERTURBATION ANALYSIS}
 We consider a compressible fluid and neglect the self-gravity, which
is ineffective in our interesting place of the Orion nebula as outlined by Bern\'{e} and Matsumoto~(2012). Our basic
equations are standard ideal MHD equations with assumption that the
gas and the radiation are in perfect radiative equilibrium. Also, we assume that there is suitable isotropic thermal coupling between matter and radiation, so that the
radiation pressure, $P_{r}$, and the radiation energy density,
$ E_r$, can approximately be related as $ P_{r}=E_{r}/3$. In this situation,
the radiation conduction approximation gives us a good relation for the diffusive flux as
(e.g., Shu~1992)

\begin{eqnarray}\label{fr}
\textbf{F}_r = - \frac{c}{3\rho\kappa} \nabla E_r = - \frac{c}{\rho\kappa} \nabla P_r
\end{eqnarray}
where $\kappa$ is the mean opacity of the gas at the interface
layer. In this way, the MHD equations in the presence of radiation field are
\begin{eqnarray}\label{10}\frac{\partial\rho}{\partial t}+\nabla.(\rho \bf{v})=0,\end{eqnarray}
\begin{eqnarray}\frac{\partial{\bf v}}{\partial t}+({\bf v}.\nabla){\bf v}=-\frac{1}{\rho}\nabla( P_{g}+ P_{r}+\frac{B^{2}}{2\mu_{0}})+\frac{1}{\rho\mu_{0}}({\bf B}.\nabla){\bf B},\end{eqnarray}
\begin{eqnarray}\frac{\partial {P}_{g}}{\partial t}=-({\bf v}.\nabla) P_{g}-\gamma P_{g}(\nabla.\bf{v}),\end{eqnarray}
\begin{eqnarray}\frac{\partial \bf{B}}{\partial t}=\nabla\times(\bf{v}\times\bf{B}),\end{eqnarray}
\begin{eqnarray}\frac{\partial E_{r}}{\partial t}+\nabla.({\bf F_{r}}+ E_{r}{\bf v})=-P_ { r}\nabla.{\bf v} \end{eqnarray}
(e.g., Shu~1992) where $ \rho $ is the fluid mass density, $ \textbf{v}$ is the
velocity, $ P_{g}$ is the gas pressure, $ B$ is the magnetic
field, and $\gamma$ is the gas adiabatic index. Considering $E_{r}=3P_{r}$ and equation (5), the equation (10) can be rewritten as:
\begin{eqnarray} \frac{\partial P_{r}}{\partial t}+\nabla.({- \frac{c}{3\rho\kappa} \nabla P_{r}}+ P_{r}{\bf v})=-\frac{P_{ r}}{3}\nabla.{\bf v}\end{eqnarray}

We perturb quantities linearly from an equilibrium state as
\begin{eqnarray*}
\rho &=&\rho_{0}+\rho_{1},\\
 \bf V&=&\bf V_{0}+ \bf V_{1},\\
P_{g}&=&P_{0g}+ P_{1g},\\
P_{r}&=&P_{0r}+ P_{1r},\\
\bf {B}&=&\bf {B_{0}}+ \bf {B_{1}}.\end{eqnarray*}
We consider the perturbations in the form of \textbf{$f_{1}(y)$}
exp$[i(k_{x}x-\omega t)]$, where \textbf{$f_{1}(y)$} is the general form
of the amplitude of perturbations of \textbf{$\rho_{1}(y)$}, \textbf{$v_{1x}(y)$}, \textbf{$v_{1y}(y)$},\textbf{$v_{1z}(y)$}, \textbf{$P_{1g}(y)$},
\textbf{$P_{1r}(y)$} ,\textbf{ $B_{1x}(y)$}, \textbf{ $B_{1y}(y)$} and \textbf{$B_{1z}(y)$}, and $k_{x}$ and $\omega$ are the wave number in the
$x$-direction and the angular frequency, respectively, so that
$\omega=\omega_{r}+i\omega_{i}$, where $\omega_{r}$ and $\omega_{i}$
are the real and imaginary parts of the angular frequency. In this way, the linearized form of the MHD
equations (6)-(7)-(8)-(9)-(11) can then be written as follows
\begin{eqnarray}
\omega \rho_{1} &=& k_{x} V_{0x} \rho_{1}+k_{x}\rho_{0} v_{1x}-i\rho_{0}\frac{\partial v_{1y}}{\partial y},\nonumber\\
\omega v_{1x}&=& k_{x} V_{0x} v_{1x}-i\frac{\partial V_{0x}}{\partial y}v_{1y}+\frac{k_{x}}{\rho_{0}\mu_{0}}P_{1g}\nonumber\\ &+&\frac{k_{x}}{\rho_{0}\mu_{0}} P_{1r}+\frac{i}{\rho_{0}\mu_{0}}\frac{\partial B_{0x}}{\partial y} B_{1y}+\frac{k_{x}}{\rho_{0}\mu_{0}} B_{0z} B_{1z},\nonumber\\
\omega v_{1y}&=& k_{x}V_{0x}v_{1y}-\frac{i}{\rho_{0}}\frac{\partial P_{1g}}{\partial y} -\frac{i}{\rho_{0}}\frac{\partial P_{1r}}{\partial y}\nonumber\\&-&\frac{i}{\rho_{0}\mu_{0}}\frac{\partial B_{0x}}{\partial y}B_{1x}-\frac{i}{\rho_{0}\mu_{0}}B_{0x}\frac{\partial B_{1x}}{\partial y}\nonumber\\&-& \frac{k_{x} B_{0x}}{\rho_{0}\mu_{0}}  B_{1y}-\frac{i}{\rho_{0}\mu_{0}}\frac{\partial B_{0z}}{\partial y} B_{1z}-\frac{i}{\rho_{0}\mu_{0}} B_{0z}\frac{\partial B_{1z}}{\delta y},\nonumber\\
\omega v_{1z}&=& k_{x} V_{0x} v_{1z}+\frac{i}{\rho_{0}\mu_{0}}\frac{\partial B_{0z}}{\partial y} B_{1y}-\frac{k_{x}B_{0x}}{\rho_{0}\mu_{0}} B_{1z},\nonumber\\
\omega P_{1g}&=& \gamma P_{0g}k_{x}v_{1x}-i\frac{\partial P_{0g}}{\partial y}v_{1y}-i\gamma P_{0g}\frac{\partial v_{1y}}{\partial y}+k_{x} V_{0x}P_{1g},\nonumber\\
\omega {P_{1r}}&=& -\frac{ic}{3\kappa}(\frac{2}{\rho_{0}^{3}}\frac{\partial \rho_{0} }{\partial y}\frac{\partial P_{0r}}{\partial y}){\rho_{1}} -\frac{i}{\rho_{0}^{2}}\frac{\partial P_{0r}}{\partial y}\frac{\partial {\rho_{1}}}{\partial y}+\frac{4}{3} P_{0r} k_{x} {v_{1x}}\nonumber\\&-& i \frac{\partial P_{0r}} {\partial y}{v_{1y}}-\frac{4}{3}i P_{0r}\frac{\partial {v_{1y}}}{\partial y}+ (k_{x} V_{0x}+\frac{ick^{2}}{3\kappa \rho_{0}}){P_{1r}}\nonumber\\&+& \frac{ic}{3\kappa \rho_{0}^{2}}\frac{\partial \rho_{0}}{\partial y}\frac{ \partial {P_{1r}}}{\partial y} -\frac{ic}{3\kappa \rho_{0}}\frac{\partial^{2}{P_{1r}}}{\partial y^2},\nonumber\\
\omega {B_{1x}}&=&-i\frac{\partial B_{0x}}{\partial y}{v_{1y}}-i B_{0x}\frac{\partial {v_{1y}}}{\partial y}+ k_{x} V_{0x}{ B_{1x}} +i\frac{\partial V_{0x}}{\partial y}{B_{1y}},\nonumber\\
\omega {B_{1y}}&=& -k_{x}B_{0x}{v_{1y}}+k_x V_{0x}{B_{1y}},\nonumber\\
\omega {B_{1z}}&=& k_{x}B_{0z}{v_{1x}}- i\frac{\partial B_{0z}}{\partial y}{v_{1y}}-iB_{0z}\frac{\partial {v_{1y}}}{\partial y}\nonumber\\&-&k_{x}B_{0x}{v_{1z}}+k_{x}V_{0x}{B_{1z}}.
\end{eqnarray}
We normalize the mass density by the density of
HII region, $\rho_{II}$, the magnetic field by the strength of magnetic field opted between two regions $B_0$, the velocity by the total variation of velocity across the layer,
$V_{0}$, the pressure $P_{0g}$, $P_{0r}$, {$P_{1g}$} and {$P_{1r}$} by $ B_{0}^{2}/2\mu_{0}$. Our analysis is done for the case in which the magnetic field is assumed to be perpendicular to the flow velocity (i.e., $B_{0x}=0$), as this case has the
fastest growth rates, according to the results of Bern\'{e} and
Matsumoto (2012). The density perturbation in the adiabatic conditions produces a pressure perturbation in accordance with $P_{g1}$= $\gamma
P_{0g}$ {$\rho_{1}$}/$\rho_{0}$  that we will consider it in the above equations. Choosing $\eta_{0}(y)=\mu_{0}V_{0}^{2}\rho_{0}(y)/{B_{0}^{2}}$ produces the normalized equations as
\begin{eqnarray}
\omega {v_{1x}}&=& k_{x} V_{0x} {v_{1x}}-i\frac{\partial V_{0x}}{\partial y}{v_{1y}}+\frac{k_{x}}{2\eta_{0}(y)}{P_{1g}}\nonumber\\&+&\frac{k_{x}}{2\eta_{0}(y)}{P_{1r}}+ \frac{k B_{0z}{B_{1z}}}{\eta_{0}(y)},\nonumber\\
\omega {v_{1y}} &=& k_{x} V_{0x} {v_{1y}}-\frac{i}{2\eta_{0}(y)}\frac{\partial {P_{1g}} }{\partial y}-\frac{i}{2\eta_{0}(y)}\frac{\partial {P_{1r}} }{\partial y}\nonumber\\&-& \frac{i}{\eta_{0}(y)}B_{0z}\frac{\partial {B_{1z}}}{\partial y},\nonumber\\
\omega {P_{1g}} &=& \gamma P_{0g}k_{x}{v_{1x}}-i\gamma
P_{0g}\frac{\partial {v_{1y}}}{\delta y}+k_{x} V_{0x}{P_{g1}},\nonumber\\
\omega {P_{1r}} &=& \frac{4}{3} P_{0r}k_{x}{v_{1x}}-i\frac{4}{3} P_{0r}\frac{\partial {v_{1y}}}{\delta y}+(k_{x} V_{0x}+ \varepsilon \frac{ick_{x}^{2}}{V_{0}}){P_{1r}}\nonumber\\&+& \varepsilon \frac{ic}{3\rho_{0}(y) V_{0}}\frac{\partial \rho_{0}}{\partial y}\frac{\partial {P_{1r}}}{\partial y}-\varepsilon \frac{ic}{3V_{0}}\frac{\partial^2 {P_{1r}}}{\partial y^2},\nonumber\\
\omega {B_{1z}} &=& k_{x}B_{0z}{v_{1x}}-iB_{0z}\frac{\partial {v_{1y}}}{\partial y}\nonumber\\&-&k_{x}B_{0x}{v_{1z}}+k_{x}V_{0x}{B_{1z}}.
\end{eqnarray}
where $\varepsilon = 1/(\kappa \rho_0(y) L)$. Here, $\tau \equiv \kappa \rho_0(y) L$ can be treated as the mean optical thickness of the gas in the transition zone . This area is considered to be opaque for input radiation, because all radiation emitted from the Trapezium, which is mostly ionizing photons ($\rm FUV$ photons), will be absorbed in the molecular cloud to ionize it. Thus, in a good approximation, the photons can not escape therefrom (mainly, we introduced the interface layer an area between fully ionized HII region and fully neutral molecular cloud), so the intensity of photons reduces to zero in the molecular cloud, $I =I _{0}e^{-\tau }\rightarrow 0$. We consider the mean optical thickness of gas in this area as $\tau \gg 1$ which results to the approximation $\varepsilon \rightarrow 0$.

 In this first analysis, for avoiding the complexity in numerical work, we have not considered the $y$ dependence for the radiation pressure and has been considered to be fixed in all over the area between two regions. The magnetic field is also considered as uniform according to (4). Hence, we opt reliable constant values for the magnetic field considering the real values of two regions of HII and molecular cloud. In these conditions, we solve equations and obtain the growth rates: imaginary part
of $\omega$ versus wave number $k_{x}$. For numerical work, we discretize the equations in y direction by the finite-difference method (e.g., Holmes et al. 2006, Matsumoto et al. 2012). We place boundaries at $y=\pm b$ far away from layer and set the boundary conditions as ${v_{1y}}=0$, $\partial {v_{1x}}/\partial y=0$, $\partial {P_{1g}}/\partial y=0$, $\partial {P_{1r}}/\partial y=0$ and $\partial {B_{1z}}/\partial y=0$ 
thereon. The boundary conditions are selected similar to Miura and Pritchett (1982). In this way, the discretized equations result an eigenvalue problem as
$$ \omega{\left(\begin{array}{cc}{v_{1x}}(1) \\
\vdots\\
{v_{1x}}(m)\\
{v_{1y}}(1) \\
\vdots\\
{v_{1y}}(m)\\
{P_{1g}}(1) \\
\vdots\\
{P_{1g}}(m)\\
{P_{1r}}(1) \\
\vdots\\
{P_{1r}}(m)\\
{B_{1z}}(1) \\
\vdots\\
{B_{1z}}(m)
\end{array}
\right)}= \widetilde{M} {\left(\begin{array}{cc}
{v_{1x}}(1) \\
\vdots\\
{v_{1x}}(m)\\
{v_{1y}}(1) \\
\vdots\\
{v_{1y}}(m)\\
{P_{1g}}(1) \\
\vdots\\
{P_{1g}}(m)\\
{P_{1r}}(1) \\
\vdots\\
{P_{1r}}(m)\\
{B_{1z}}(1) \\
\vdots\\
{B_{1z}}(m)
\end{array}
\right)}$$
where $m$ is the number of grid points. We
obtain the eigenvalues of the matrix $\widetilde{M}$ and consider the imagenary part of $\omega$ which is the growth rate of instability. For the initial parameters, we use the values of typical densities and temperatures of the HII regions and the molecular clouds in the ISM (Tielens 2005), observational constraints which were utilized by Matsumoto and Bern\'{e} (2012), and the estimated values of the radiation pressure of Trapezium that are obtained in the section~2.
\section{Results}
\subsection{Importance of the radiation pressure}
By solving the eigenvalue problem, which was introduced in last section, results of the linear analysis, concerned with the linear growth of the KH instability, are presented in this section. Figure 4 shows the normalized growth rate of instability  $\omega_{i}L/V_{0}$ as a function of normalized wave number $k _{x}L $, for the sample of five selected values of the ratio of radiation pressure to gas pressure,  $P_{0r}/P_{0g}$, introduced by $\Pi$. Each curve demonstrates different value of $\Pi$, from bottom to top as 0, 10, 30, 50, 100. The different values for $\Pi$ are selected according to some of possible values of the ratio $P_{0r}=xL^{T} /4\pi c d^{2}$ to $P_{0g}= k_B n_{ II} T_{ II} = k_B n_{ MC} T_{ MC}$ where coefficient $x$ shows the fraction of emitted photons which are able to enter to the interface layer. It varies between 0 to 1 depending on amount of absorption and scattering by the HII region. In this medium, we assume that the initial values for number densities and temperatures of two regions of the HII and molecular cloud are as $n_{II}=20\rm cm^{-3}$, $T_{ II}=10^{4}\rm K$ and  $n_{MC}=10^{4}\rm cm^{-3}$, $T_{MC}=20\rm K$, respectively. With these typical values of physical quantities and typical distances of 2pc and 3pc, we can find reliable values for the importance of radiation pressure which are obtained as $0 <\Pi< 100.$ Initial values for the magnetic field strength, and the variation of the velocity between two regions, are considered as 20nT and $ 10 \rm km s^{-1} $, respectively. 

The lowermost curve in the Figure 4, represents the case in which there is no radiation pressure. Clearly for $\Pi=0$ we can find the same solution  presented by Matsumoto and Bern\'{e}(2012). As $\Pi$ increases from zero (and hence radiation pressure becomes important), the normalized growth rate of KH modes increases. In addition, in the presence of the radiation pressure, the domain of instability is extended as far as it includes more wavelengths consisting of smaller ones rather than the case when the radiation pressure is disregarded. On the other hand for $k_{x}L < 0.4(\lambda _{\rm KH}>0.15 \rm pc),$ the growth rate is not sensitive to radiation pressure.
 
Figure 5 shows the influence of the different values of $\Pi$ on the normalized most unstable modes $(\omega_{i}L/V_{0})_{\rm max} $ [left] and the normalized wave numbers of most unstable modes $ (k_{x}L)_{\rm max}$ [right]. It clearly shows that the most unstable modes and wave numbers of most unstable modes are shifted toward the large values under influence of the radiation pressure. Meaning that most unstable wavelengths, $\lambda _{\rm KH} $ are shifted toward small ones, implying that the more unstable wavelengths are small wavelengths. It is expected that in the regions which are subject to high radiation pressure, structures with small wavelengths to be occurred more.
\begin{figure}
\includegraphics[width=7cm,height=4.5cm]{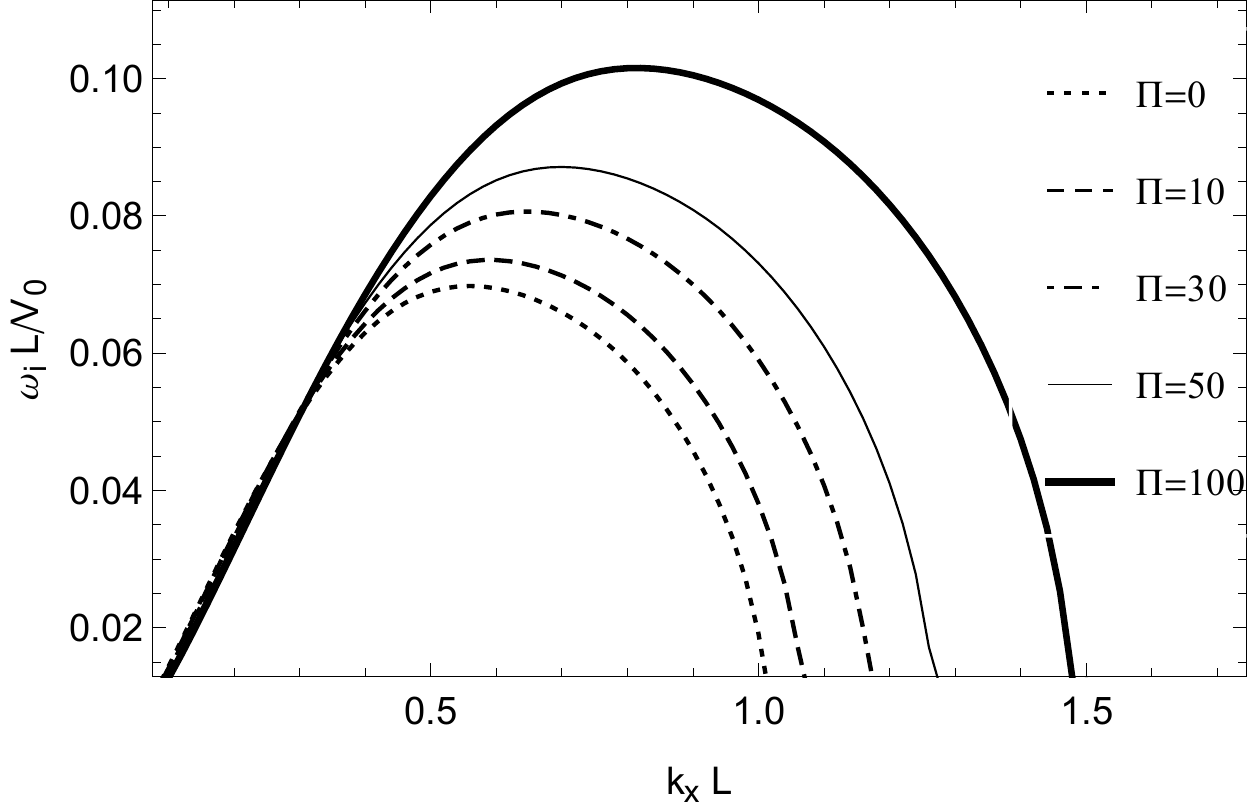}
\caption{The normalized growth rates of instability $ \omega_{i}L/V_{0}$ as a function of normalized wave number $k _{x}L $. Each curve demonstrates different value for $\Pi=P_{0r}/P_{0g}$ from bottom to top as 0, 10, 30, 50, 100. The lowermost curve represents the case in which there is no radiation pressure. Magnetic field strength, total variation of the velocity and density contrasts between two regions are considered as $ 20\rm nT$, $ 10 \rm km s^{-1} $ and $ 500 $, respectively. Number densities and temperatures are taken to be $n_{II}=20 \rm cm^{-3}$, $T_{ II}=10^{4}\rm K$ and $n_{MC}=10^{4}\rm cm^{-3}$, $T_{MC}=20\rm K$ . \label{???}}
\end{figure}

Figure 6 [top] shows e-fold time-scales of instability as a function of wave numbers $k_ {x}L$ for different values of $\Pi$. Each curve demonstrates different values for $\Pi$, from top to bottom as 0, 10, 30, 50, 100.
\begin{figure}
\includegraphics[width=4.1cm,height=4.3cm]{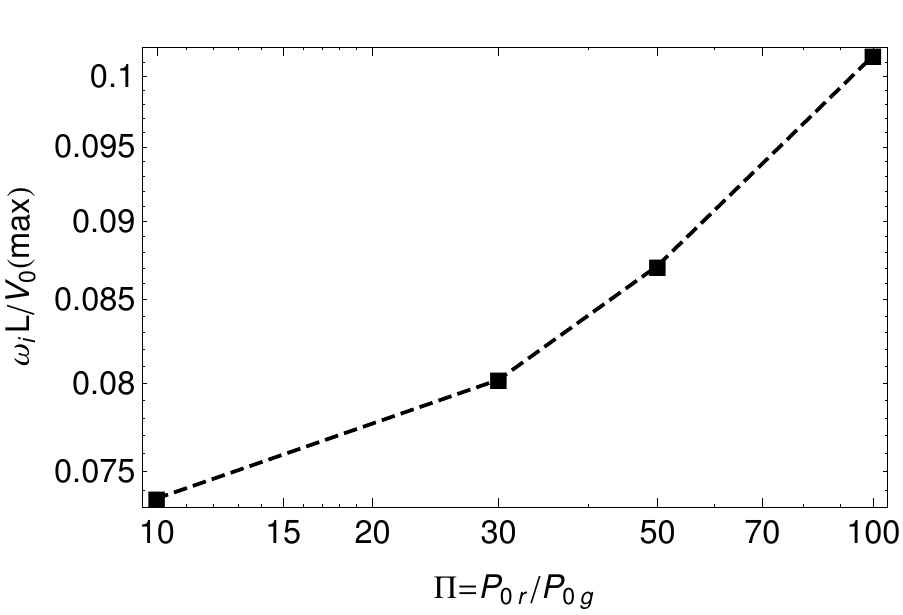}
\includegraphics[width=4.1cm,height=4.3cm]{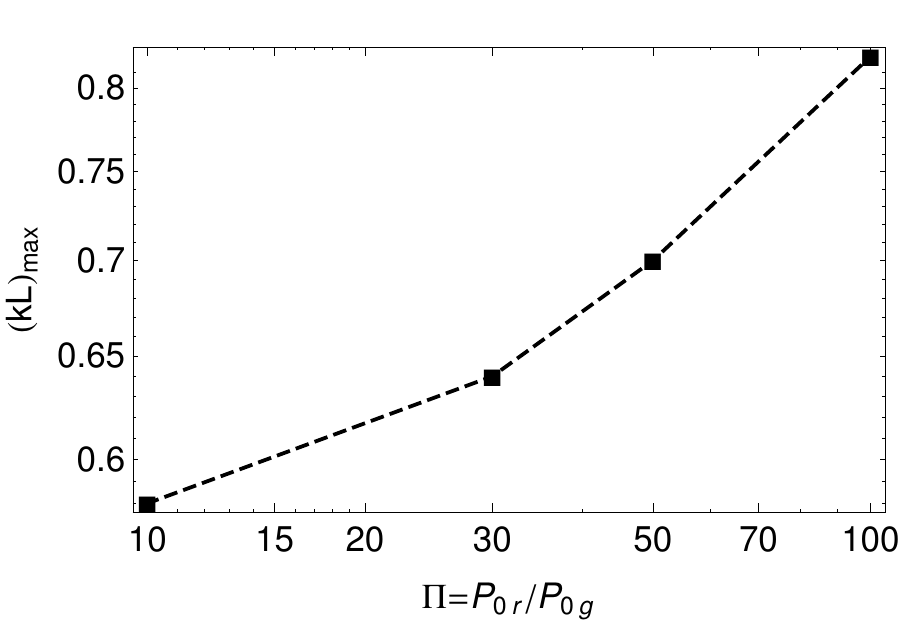}
\caption{Left: The normalized growth rate of the most unstable modes of instability as a function of $\Pi=P_{0r}/P_{0g}$. Right: The normalized wave number of most unstable modes as a function of $\Pi=P_{0r}/P_{0g}$. \label{???}}
\end{figure} 
The uppermost curve in figure, represents the case in which there is no radiation pressure. The e-fold time-scale decreases by increasing the radiation pressure, while e-fold time-scale of the instability for $k_{x}L < 0.4 $ does not depend to the radiation pressure. This figure also shows that in the large wave numbers, reduction rate of e-fold time-scales is more than other wave numbers.
Figure 6([bottom] illustrates the e-folds time-scales versus the $\Pi$ for typical large wave numbers $ k_{x}L=0.8$, $ k_{x}L=0.9$ and $ k_{x}L=1$. By increasing the radiation pressure from 0 to $ \sim 100 $ times of gas pressure, e-fold time-scale reduces from $ 1.7\times 10^{4} $ year, $ 2.2\times 10^{4} $ year and $ 5\times 10^{4} $ year to about $ 10^{4}$ year for $k _{x}L=0.8$, $k _{x}L=0.9$ and $k _{x}L=1$, respectively. Meaning that small wavelengths develop faster than other wavelengths in the domain $k_{x}L >0.4(\lambda _{\rm KH}<0.15\rm pc)$. 

\begin{figure}
\includegraphics[width=7cm,height=4.5cm]{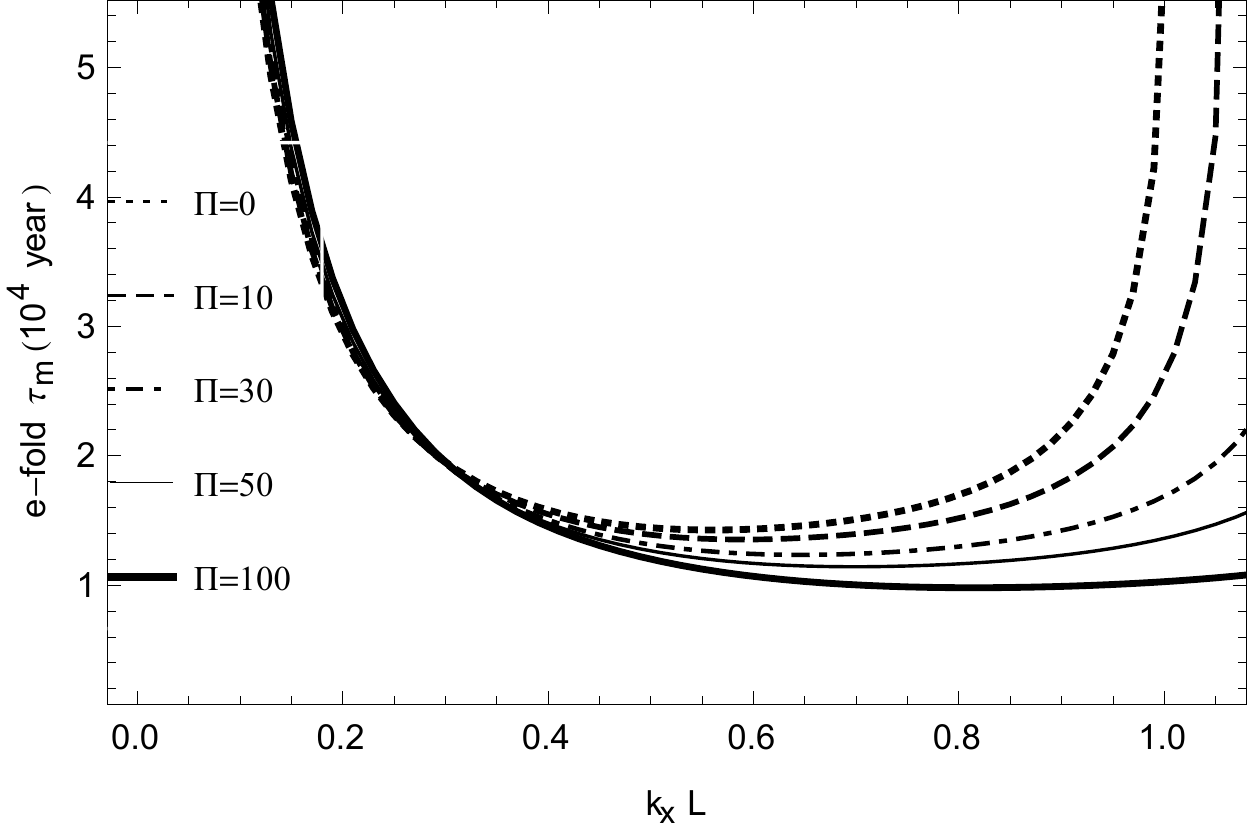}
\includegraphics[width=7cm,height=4.5cm]{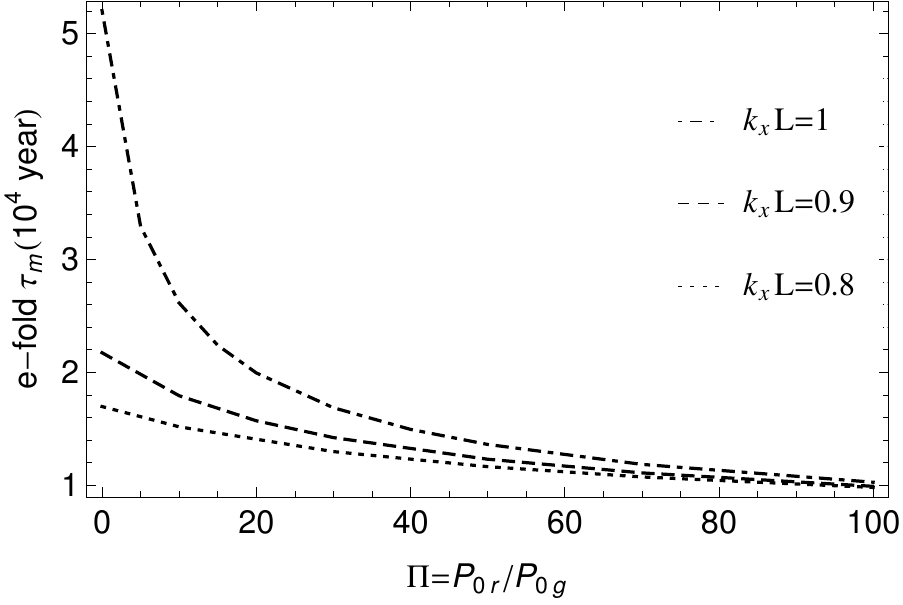}
\caption{Top: The e-fold time-scales of the KH instability as a function of the normalized wave number. Each curve indicates different value for $\Pi=P_{0r}/P_{0g}$ from top to bottom as 0, 10, 30, 50, 100. The uppermost curve represents the case in which the radiation pressure is absent. Bottom: The e-fold time-scale of the KH instability as a function of $\Pi=P_{0r}/P_{0g}$ for $ k_{x}L=0.8$, $ k_{x}L=0.9$ and $ k_{x}L=1$.\label{???}}
\end{figure}
\subsection{Importance of the magnetic field strength} 
It is now accepted that the magnetic fields play a significant role on the dynamics of the ISM, and consequently on the star forming regions of the Orion nebula (e.g., Abel et al. 2004; Brogan et al. 2005; Heiles et al. 1993; McKee et al. 1993; Dib et al. 2010; Nejad-Asghar 2016; Hosseinirad et al. 2017). It makes Orion nebula a proper lab for studying of the magnetic fields and their effects on the surrounding nebula, specially the instabilities formed through the clouds. Since magnetic field is an important source of energy and pressure in the Orion nebula, it worth to study the influences of magnetic field strength on the evolution of KH instabilities. 
\begin{figure}
\includegraphics[width=6.5cm,height=4.5cm]{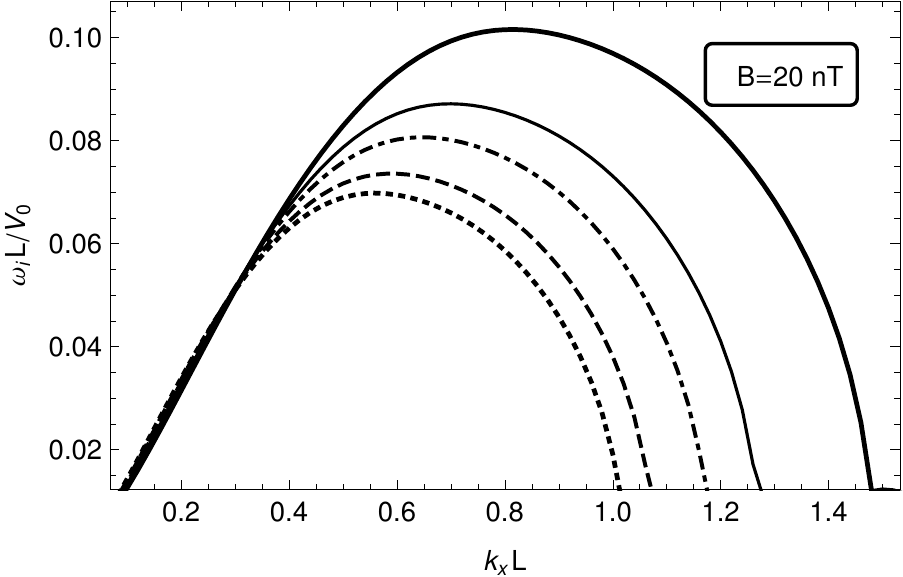}
\includegraphics[width=6.5cm,height=4.5cm]{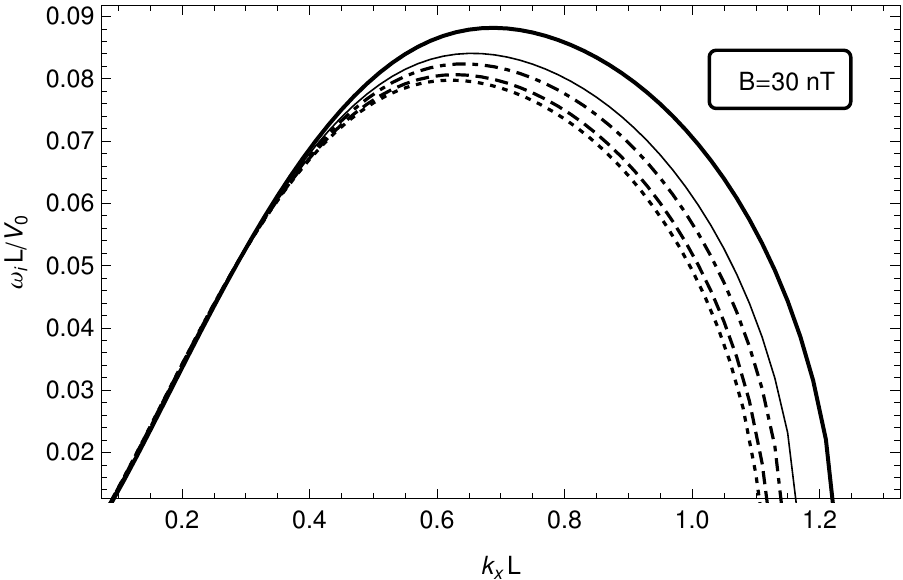}
\includegraphics[width=6.5cm,height=4.5cm]{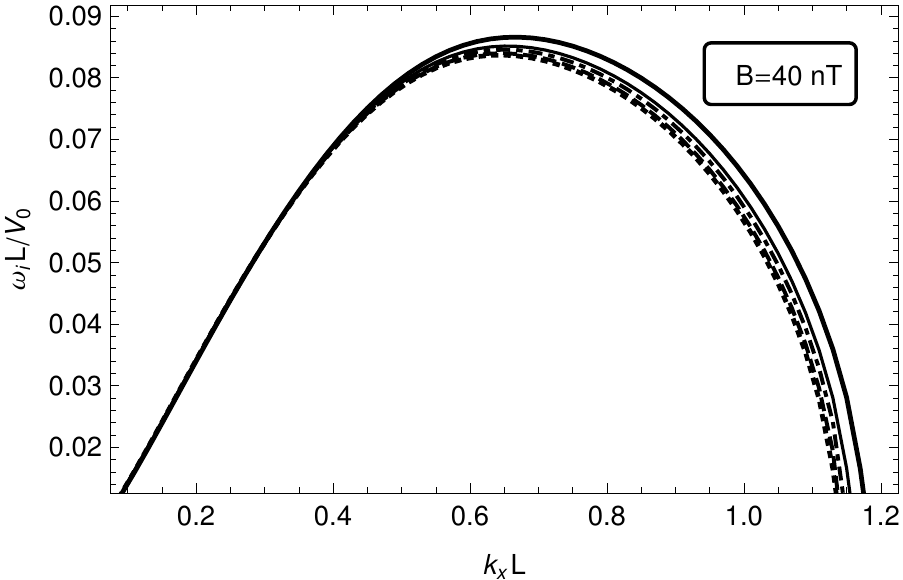}
\caption{The normalized growth rates of the KH instability as a function of normalized wave numbers in the presence of the radiation pressure, like as Figure 4, but for different values of magnetic fields strength 20nT, 30nT and 40nT. \label{???}}
\end{figure}
Figure 7 shows the growth rate of instability versus the normalized wave number in the presence of the radiation pressure for three different magnetic field strengths: 20nT, 30nT and 40nT, in upper, middle and lower panels, respectively. Our investigation here on the variation of the magnetic field strength is motivated mainly by the well-known model which pointed out that there is a correlation between magnetic field strength and density of ISM as $ B \propto n^{a}$ with $ a\sim 0.5$ (Crutcher~2012). According to it, denser molecular clouds contain the stronger magnetic fields. Here, we adopt three values for the magnetic field strength 20nT, 30nT and 40nT which are correspond to $ n_{MC}=10^{4}\rm cm^{-3}$, $ n_{MC}=2\times10^{4}\rm cm^{-3}$ and $ n_{MC}=4\times10^{4}\rm cm^{-3}$ respectively. The gas pressure is considered to be fixed for three cases mentioned above. In these conditions, $P_g= nkT= const.$ implies that the denser molecular clouds are assumed to have lower temperatures.

 As Figure 7 clearly shows, by increasing the magnetic field strength, from top to bottom, the increasing effects of the KH instabilities induced by radiation pressure will suppress. As we can deduce from this figures, the radiation pressure enhances the growth rate, while the magnetic field suppresses it, gradually. In the case $40nT$, the growth rate almost dose not depend to the radiation pressure. 

It can be interpreted such that there is a competition between the radiation pressure and the magnetic pressure $P_{0m}= B_{0}^{2}/2\mu_{0}$. By setting the radiation pressure as $P_{0r}= 10-100 P_ {0g}$ and the magnetic field strength as $B _{0}=20\rm nT $, we obtain $P_{0r}/P_{0m} = 0.17-0.87 $. As the magnetic field increases, for the magnetic field $ B_{0}=30\rm nT$ and $ B_{0}=\rm 40nT$, $P_{0r}/P_{0m}$ gives $0.076-0.384$ and $0.043-0.216$, respectively. Meaning that by increasing the magnetic pressure, the contribution of radiation pressure decreases and the magnetic pressure is dominant over the radiation pressure. In these cases of high magnetic fields, the growth rate of instability is independent of radiation pressure and depends only on the magnetic pressure. In the other words, the magnetic field play a controller role in development of the KH instabilities.

\section{DISCUSSION}
We studied the dynamical effects of radiation pressure, emerged from the Trapezium cluster, on the growth rate of KH instabilities in the Orion nebula, which is representative of many star forming regions. The obtained results can be used for other star formation regions, which have the favorable conditions for forming rippled structures.
The results show that the radiation pressure has a intensifying effect on the growth rate of the KH modes and decreases e-fold time-scale of instability. The domain of instability is also extended and includes the more wavelengths including smaller ones rather than the cases when the effect of the radiation pressure is disregarded. It is also resulted that for $\lambda_{\rm KH}>0.15\rm pc$, the growth rate of instability do not influenced by the radiation pressure. Considering that the most unstable wavelengths are shifted toward smaller values, it is expected that, in the regions which are subjected to the high radiation pressure, the KH instabilities with the small scales to be occurred  more likely.
Mutsumoto and Bern\'{e} (2012) studied the evolution of linear phase of KH instability toward saturation phase and claimed that the turbulent structures are formed during time $t_{\rm sat} $. Our results show that the radiation pressure accelerates the development of KH instabilities in the linear phase and subsequently, it is expected that the formation of turbulent structures advances. Since e-fold time-scales of instability are short comparing to the life-time of the OB associations ($ \sim 10\rm Myr $), thus the radiation pressure can act effectively on its surrounding. As well as, the radiation pressure is probably a significant factor to generate small scale turbulences in molecular clouds near massive stars. These results suggest that radiation pressure is a triggering factor in the structure of ISM.

The role of magnetic field strength in the presence of the radiation pressure is also discussed. As the strength of the magnetic field increases, the growth rate of instability becomes independent of radiation pressure and it depends only on the magnetic pressure. In the other word, the increasing effects of instabilities induced by radiation pressure will suppressed by the magnetic field, meaning that magnetic field plays a controller role for evolution of KH instabilities induced by the radiation pressure.  
Given these results, it remains to future work to extend this analysis to full implications of radiation pressure, non-ionizing radiation, non-ideal MHD case, and the effects concerned with dust.

This work was supported by Ferdowsi University of Mashhad under grant 3/41769 (1395/06/31). We also appreciate the referee for his/her thoughtful and constructive comments on an early version of the paper. We also thanks Mahmood Roshan for his useful discussions.

\end{document}